%% file: fmas2020_main.tex
\documentclass[submission,copyright,creativecommons]{eptcs}
\providecommand{\event}{FMAS 2020} 
\usepackage{breakurl}             
\usepackage{underscore}           
\usepackage{bsymb,newestb2latex}
\usepackage{programs}

\usepackage[xindy]{glossaries}
\loadglsentries{acronyms}

\usepackage{tabularx}

\usepackage{booktabs} 
\usepackage{amsmath,amssymb,amsfonts}
\usepackage{algorithmic}
\usepackage{graphicx}
\usepackage{textcomp}
\usepackage{xcolor}
\usepackage{url}
\usepackage{pifont}
\newcommand{\xmark}{\ding{55}}%
\usepackage{tabularx}

\usepackage{tikz}
\usetikzlibrary{shapes.geometric, arrows, shapes,backgrounds,arrows,calc,positioning}
\tikzstyle{process} = [rectangle, minimum width=2.5em, minimum height=2em, text centered, draw=black]
\tikzstyle{arrow} = [thick,->,>=stealth]
\usepackage{tablefootnote}

\usepackage{xspace}
\usepackage{adjustbox}

\usepackage{zed-csp}
\usepackage{gensymb}

\usepackage{regexpatch}
\usepackage[colorinlistoftodos]{todonotes}

\makeatletter
\xpatchcmd{\@todo}{\setkeys{todonotes}{#1}}{\setkeys{todonotes}{inline,#1}}{}{}
\makeatother

\title{
Towards Compositional Verification \\ for Modular Robotic Systems
\thanks{Work supported by UK Research and Innovation, and EPSRC Hubs for Robotics and AI in Hazardous Environments: EP/R026092 (FAIR-SPACE), EP/R026173 (ORCA), and EP/R026084 (RAIN).}
}

\author{
    Rafael C. Cardoso \quad Louise A. Dennis \quad Marie Farrell  \quad Michael Fisher  \quad Matt Luckcuck
    \institute{Department of Computer Science\\ The University of Manchester\\ Manchester, United Kingdom}
    \email{\{rafael.cardoso, louise.dennis, marie.farrell, michael.fisher, matthew.luckcuck\}@manchester.ac.uk}
}

\def\titlerunning{Towards Compositional Verification for Modular Systems}
\def\authorrunning{R.C. Cardoso et al.}

\begin{document}
\maketitle

\begin{abstract}
Software engineering of modular robotic systems is a challenging task, however, verifying that the developed components all behave as they should individually and as a whole presents its own unique set of challenges. In particular, distinct components in a modular robotic system often require different verification techniques to ensure that they behave as expected. Ensuring whole system consistency when individual components are verified using a variety of techniques and formalisms is difficult. This paper discusses how to use compositional verification to integrate the various verification techniques that are applied to modular robotic software, using a First-Order Logic (FOL) contract that captures each component's assumptions and guarantees. These contracts can then be used to guide the verification of the individual components, be it by testing or the use of a formal method. 
We provide an illustrative example of an autonomous robot used in remote inspection. We also discuss a way of defining confidence for the verification associated with each component.
\end{abstract}



\section{Introduction}

Autonomous robotic systems are being used more frequently in safety-critical scenarios. Examples include monitoring offshore structures~\cite{Hastie18}, nuclear inspection and decommissioning~\cite{Bogue:11}, and space exploration~\cite{Wilcox:92}. Ensuring that the software which controls the robot behaves as it should is crucial, particularly as modern robotic systems become more modular, and are deployed alongside humans in both safety- and mission-critical scenarios.

Robotic software is often developed using a middleware to facilitate interoperability, such as ROS\footnote{\url{http://www.ros.org/}} or \Genom{}\footnote{\url{https://www.openrobots.org/wiki/genom}}. They share some abstract concepts~\cite{Shakhimardanov2010}, namely that systems are composed of communicating components. The approach that we describe in this paper is not restricted to any particular robotic middleware, instead it can be applied to a range of modular systems.

Distinct components in a modular system often require different verification techniques, ranging from software testing to formal methods. In fact, \emph{integrating} (formal and non-formal) verification techniques is crucial particularly for robotic applications~\cite{Farrell2018}. It is essential for the verification to be carried out using the most suitable technique or formalism for each component. However, performing compositional verification via linking heterogeneous verification results of individual components is difficult and the current state-of-the-art for robotic software development does not provide an easy way of achieving this. 

Our approach is to construct a high-level \gls{fol} contract specification of the system. The \gls{fol} contract describes the expected input, required output, and Assume-Guarantee~\cite{Jones83} conditions for each component. This abstract specification can be seen as a logical prototype for individual components and, via our calculus for chaining individual specifications, the entire modular system. Once the \gls{fol} contracts have been checked for consistency, they can be used to guide the verification of each component. This involves ensuring that the verification encodes the assumptions and guarantees as, for example, test cases, assertions, or formal properties.

Our approach can be used in a Top-Down manner, to guide the system's development from abstract specification to concrete
implementation, via verification; or in a Bottom-Up fashion, to check the  consistency of existing verification techniques. In this paper we present a Top-Down example. Our approach enables developers to choose the most suitable verification technique for each component, but also to link the conditions being verified across the whole system.

The next section presents background and related work. Section~\ref{sec:specifyingRobots} describes our contribution for specifying modular systems using contracts written in \gls{fol}; then, we illustrate our approach via a simple remote inspection example. In Section~\ref{sec:confidence} we discuss how the use of distinct forms of verification (e.g., testing, simulation, formal methods, etc.) affects our confidence in the verification results for the whole system. Finally, Section~\ref{sec:conclusion} concludes and outlines future directions.

\section{Background and Related Work}
\label{sec:related}

Modular robotic systems that are used in safety- or mission-critical scenarios require robust verification techniques to ensure and certify that they behave as intended. These techniques encompass a range of software engineering tools and methodologies: from practical testing and simulation, through to formal verification~\cite{Luckcuck2019,Farrell2018}. 
For example, to show that a property always holds: model-checkers~\cite{clarke1999model} exhaustively explore the search space, and theorem provers~\cite{bertot2013interactive} use mathematical proof. These techniques offer a means for proving that the software system is correct, which can be used as evidence to improve public trust or to gain certification, as needed. Formal verification can be applied to the implemented system; or to an abstract specification, which can then be further refined to program code.

Many formal methods exist, and most follow a simple paradigm: if the program is executed in a state satisfying a given property (the pre-condition), then it will terminate in a state that satisfies another property (the post-condition)~\cite{dijkstra1975guarded}. It is up to the software engineer to specify these properties (sometimes referred to as assumptions and guarantees, respectively). The \emph{Assume-Guarantee} (or Rely-Guarantee) approach~\cite{Jones83} enables components to be assessed (verified) individually and then the separate assessments to be combined, potentially under some assumption about the underlying concurrency. To combine assessments, specifications of a component's pre- and post-conditions are needed; this is similar to the notion of Hoare logic that underlies these techniques~\cite{hoare1969axiomatic}. In this work, we refer to this combination of assumption/pre-condition and guarantee/post-condition as a \textit{contract} that must be preserved.
 

Our work takes inspiration from Broy's logical approach to systems engineering~\cite{Broy18, Broy18b} in which he distinguishes three kinds of artefacts: (1) system-level requirements, (2) functional system specification, and (3) logical subsystem architecture. Each of these artefacts is represented as a logical predicate in the form of assertions, with relationships defined between them which he then extends to assume/commitment contracts. His treatment of these contracts is purely logical \cite{Broy18b}, and, in this paper, we present a similar technique that is specialised to the software engineering of modular robotic systems.
    

In~\cite{LuckenederK18}, a workflow is proposed for systematically verifying the design of models of a cyber-physical system using a combination of formal refinement and model-checking. 
Our work also deals with several different levels of abstraction, but we tackle the use of compositional verification in modular systems.

Recent work has analysed a portion of the literature and identified common patterns that appear in robotic missions (e.g. patrolling and obstacle avoidance)~\cite{menghi2019specification}. They provide verified LTL/CTL specifications for these commonly found missions which can then be reused in future developments.  Further, it is not clear how their support for compositional verification can be extended to support heterogeneous components such as those in our example. Related work includes the identification of Event-B specification clones in cyber physical system specifications~\cite{farrell2017specification}. 

Other compositional approaches include OCRA~\cite{cimatti2013ocra}, AGREE~\cite{cofer2012compositional}, CoCoSpec~\cite{champion2016cocospec}, SIMPAL~\cite{SIMPAL}, DRONA~\cite{DRONA}, and a methodology to decompose a system into assume-guarantee contracts that are then validated through simulation~\cite{Spellini2019}, however, none explicitly incorporates heterogeneous techniques.

\section{A First-Order Logic Framework}
\label{sec:specifyingRobots}
\label{sec:spec}

No single verification approach suits every component in a modular robotic system~\cite{Farrell2018}. Components such as hardware interfaces or planners may be amenable to formal verification, whereas, sensors may require software testing or simulation-based testing.

As illustrated in Fig.~\ref{fig:nodespecs}, we could use logical specifications (e.g. temporal logic), model-based specifications (e.g. Event-B \cite{abrialmodeling2010} or Z \cite{spivey1988understanding}), or algebraic specifications (e.g. CSP \cite{Hoare1978} or CASL\cite{mosses2004casl}) amongst others to specify the components in the system. Each of these formalisms offers its own range of benefits, and each tends to suit the verification of particular types of behaviour. Also, we may only have access to the black-box or white-box implementation of a component. 

\input{dia/nodesAndVerification}

Our approach facilitates the use of compositional verification techniques for the components in modular robotic systems. We achieve this by specifying high-level contracts in \gls{fol} and we employ temporal logic for reasoning about the combination of these contracts. In this way, we attach the assumptions over the input ($\mathcal{A}(\overline{i})$) and guarantees over the output ($\mathcal{G}(\overline{o})$) to individual components (see Fig. \ref{fig:nodespecs}).



\subsection{A Calculus for Chaining Component Specifications}
\label{sec:chainingsubsec}

Specifically, we use \emph{typed} \gls{fol}, potentially with the addition of algebraic operators, to specify assumptions and guarantees. For
each component, $C$, we specify $\mathcal{A}_C(\overline{i}_C)$ and $\mathcal{G}_C(\overline{o}_C)$ where $\overline{i}_C$ is a variable representing the input to the component, $\overline{o}_C$ is a variable representing the output from the component, and $\mathcal{A}_C(\overline{i}_C)$ and $\mathcal{G}_C(\overline{o}_C)$ are \gls{fol} formulae describing the assumptions and guarantees, respectively.

Each individual component, $C$, obeys the following implication:\\
\centerline{$\forall \overline{i}_C, \overline{o}_C \cdot  \mathcal{A}_C(\overline{i}_C)[C] \Rightarrow\ \lozenge \mathcal{G}_C(\overline{o}_C)$}

\noindent where $\overline{i}_C$ and $\overline{o}_C$ are the inputs and outputs, respectively, of $C$; $\mathcal{A}_C(\overline{i}_C)[C]$ represents the execution of $C$ with the assumption $\mathcal{A}_C(\overline{i}_C)$; and `$\lozenge$' is \gls{ltl}'s~\cite{Pnueli77temporal} ``eventually'' operator. So, this implication means that if the assumptions, $\mathcal{A}_C(\overline{i}_C)$, hold in the specification or program code of $C$, then upon execution of the component \textit{eventually} the guarantee, $\mathcal{G}_C(\overline{o}_C)$, will hold. Note that our use of temporal operators here is motivated by the real-time nature of robotic systems and could be of use in later extensions of this calculus for larger, more complex systems.

Components in a modular robotic architecture are `chained'
together so long as their types/requirements match. Similarly, we can
compose the contracts of individual components in a number of ways, the simplest being
a sequential composition. The basic way to describe
such structures is to first have the specification capture \emph{all} of
the input and output streams and then to describe how these are
combined in the appropriate inference rules. 
The proof rule, PR1, for such linkage is as expected:
$$
\begin{array}{l}
\forall \overline{i}_1,\overline{o}_1.\ \mathcal{A}_1(\overline{i}_1)\ \Rightarrow\ \lozenge \mathcal{G}_1(\overline{o}_1)\\
\forall\overline{i}_2,\overline{o}_2.\ \mathcal{A}_2(\overline{i}_2)\ \Rightarrow\ \lozenge \mathcal{G}_2(\overline{o}_2)\\
\overline{o_1} = \overline{i}_2\\
\vdash\ \forall\overline{i}_2,\overline{o}_1.\ \mathcal{G}_1(\overline{o}_1)\ \Rightarrow\ \mathcal{A}_2(\overline{i}_2)\\ \hline
\forall\overline{i}_1,\overline{o}_2.\ \mathcal{A}_1(\overline{i}_1)\ \Rightarrow\ \lozenge\mathcal{G}_2(\overline{o}_2)
\end{array}
\hspace{30pt} \text{(PR1)}
$$
Intuitively, this states that if two components are sequentially composed with the output of the first equal to the input of the second and the guarantee of the first implies the assumption of the second then, we can deduce that the assumption of the first component implies that the guarantee of the second will ``eventually'' hold.

The illustrative example that we present in the next section contains a neat, linear chain of components and so it is easy to see how this proof rule works. But a realistic robotic system could be much more complex than this. 
For example, a component might have multiple, branching output streams that are each used as input to other distinct components. Thus, we propose PR2 to account for branching: 
$$
\begin{array}{l}
\forall \overline{i}_1,\overline{o}_1.\ \mathcal{A}_1(\overline{i}_1)\ \Rightarrow\ \lozenge \mathcal{G}_1(\overline{o}_1)\\
\forall\overline{i}_2,\overline{o}_2.\ \mathcal{A}_2(\overline{i}_2)\ \Rightarrow\ \lozenge \mathcal{G}_2(\overline{o}_2)\\
\overline{i}_2 \subseteq \overline{o}_1\\
\vdash\ \forall\overline{i}_2,\overline{o}_1.\ \mathcal{G}_1(\overline{o}_1)\ \Rightarrow\ \mathcal{A}_2(\overline{i}_2)\\ \hline
\forall\overline{i}_1,\overline{o}_2.\ \mathcal{A}_1(\overline{i}_1)\ \Rightarrow\ \lozenge\mathcal{G}_2(\overline{o}_2)
\end{array}
\hspace{30pt} \text{(PR2)}
$$
Here, the input to the second component, $\overline{i}_2$, is a subset of the output, $\overline{o}_1$, of the first.
We have also devised other proof rules which are omitted here for brevity.

\subsection{Remote Inspection Case Study}
\label{sec:systemSpec}

\input{dia/robotSystem}

The various middleware often used to develop robotic systems enable the combination of subsystems, potentially developed independently of one another. For example, an autonomous rover might have a navigation subsystem that is responsible for traversing a planet's surface, and a subsystem for sampling and analysing rock and soil samples.
 
In this paper, we use the illustrative example of a simplified navigation subsystem for an autonomous rover, whose goal is to perform remote inspection of particular targets, while avoiding any obstacles, on a 2D grid map that has been previously generated. This navigation subsystem is composed of components for \emph{Detection}, a \emph{Planner} and a cognitive \emph{Agent}. Fig.~\ref{fig:system} illustrates this subsystem and outlines the abstract \gls{fol} contract of each component to be verified. 

Given the camera input and the size of the square grid to be explored, $n$, the specification of the \emph{Detection} component guarantees that it detects an obstacle at a particular coordinate if an obstacle actually exists (in the physical or simulated environment) at that point. Note that we assume the existence of the $obstacle(x,y)$ function to represent this physical or simulated observation. Further, the \emph{Detection} component outputs the rover's current location, denoted by $s_0$, and guarantees that $s_0$ is in the grid and does not coincide with an obstacle.

The \emph{Planner} component produces a $PlanSet$ whose elements are sets of coordinates capturing obstacle-free plans under the assumption that the \emph{Detection} component has produced a set of $Obstacles$ that refers to points that exist and contain an obstacle. The guarantee also ensures that each of these plans (sets of coordinates) in $PlanSet$ can be transformed into a sequence of coordinates where each is adjacent to the previous. Note that we use sets here rather than sequences because sequences may contain duplicate entries whereas sets cannot, and so our specification rules out the case where a plan loops.

The specification of the cognitive \emph{Agent} component states that, under the assumption that all potential plans in the $PlanSet$ are obstacle-free, then it chooses a $plan$ from the $PlanSet$ and that this plan is, in fact, the shortest one available. Note that we use $|plan|$ to denote the length of $plan$.


The far right of Fig.~\ref{fig:system} contains an Event-B event specification~\cite{abrialmodeling2010} of the \emph{Planner} component. Notice that the high-level FOL contract that we have written is captured in the Event-B specification as guards on lines 5--7. The FOL contract has been refined and the Event-B specification has a notion of a goal that it must plan paths towards. This has resulted in the extra guards on lines 1 and 8. This is an excerpt from a larger \emph{Planner} specification that we have verified via theorem proving in the \emph{Rodin} Platform \cite{abrial2010rodin}.

In this illustrative example, we formally verified the \emph{Planner} contracts using Event-B and the \emph{Agent} contracts using the Gwendolen agent programming language \cite{Dennis2012}. The \emph{Detection} component would be verified via standard software testing practices and simulation-based testing since it may contain a learning component which cannot be formally verified. 

In \S\ref{sec:chainingsubsec}, we presented rules about how we chain contracts together. In this example, two applications of PR1 would allow us to derive the following property: \\
\centerline{$\forall \overline{i}_D, \overline{o}_A \cdot  \mathcal{A}_C(\overline{i}_D) \Rightarrow\ \lozenge \mathcal{G}_A(\overline{o}_A)$}
\noindent meaning that if the \emph{Detection} component receives valid input then the \emph{Agent} will eventually return the shortest path from the start to the goal position in the grid. 

We have shown how FOL can be used as a unifying language for the high-level Assume-Guarantee specification of components in a modular robotic system. A top-down approach to software engineering of such a robotic system would begin with these FOL contracts. Then, each contract would be further refined~\cite{morganrefinement1988}, ideally to a more detailed formal specification and, eventually to its concrete implementation.

\section{Confidence in Verification}
\label{sec:confidence}

When linking the results from multiple verification techniques a  key question becomes how using these different techniques affects our confidence in the verification of the whole system. One might think that a formal proof of correctness corresponds to a higher level of confidence than simple testing methods (especially over unbounded environments). However, formal verification is usually only feasible on an abstraction of the system whereas testing can be carried out on the implemented code. Therefore, it is our view that we achieve higher levels of confidence in verification when multiple verification methods have been employed for each component in the system \cite{WWADEMP19}. 

\begin{table}[ht]
\centering
\scalebox{0.75}{
\begin{tabular}{|m{4.3cm}|m{4.3cm}|m{4.3cm}|m{4.3cm}|}
\hline
\textbf{Component} & \textbf{Testing} & \textbf{Simulation-Based Testing} & \textbf{Formal Methods} \\
\hline \hline
\emph{Detection} & \checkmark & \xmark& \xmark \\
\hline
\emph{Planner}  &\checkmark & \checkmark& \checkmark \\
\hline
\emph{Cognitive Agent} &\checkmark & \checkmark& \checkmark \\\hline
\end{tabular}}
\caption{\small Verification techniques applied to each component.}
\label{table:confidence}
\end{table}

We have broadly partitioned current verification techniques into three categories: testing, simulation-based testing and formal methods. We have determined which of these techniques might be employed for each component in our example as shown in Table \ref{table:confidence}. 
Examining how this metric can be calculated for more complex systems with loops is a future direction for this work.




\section{Conclusions and Future Work}
\label{sec:conclusion}
We have presented an initial description of a framework for compositional verification of modular systems using \gls{fol} as a unifying language. \gls{fol} contracts are used to guide the verification techniques applied to each component. We have briefly illustrated this by verifying and refining the FOL contract for the \emph{Planner} component using Event-B in our example of a remote inspection rover. Furthermore, we introduce the notion of confidence in verification techniques and provide a broad categorisation. 


Future work includes assessing how well our approach scales to industrial-sized robotic systems development.
Also, ensuring that our approach is usable by developers and understandable by certification organisations is crucial to its use and success. This needs tool support for writing and reasoning about the \gls{fol} contracts, with similar functionality to an IDE, integrated robotic development other tools. We will also demonstrate the approach with a wider variety of verification techniques. Our illustrative example employs an event-based style of communication, however, we intend to explore how to extend our approach to stream-based communications as future work.

Finally, our proposed method could be used to generate runtime monitors to check that each component's contract holds during execution. If a contract is violated, the monitor could flag this to the robot's operator (if one exists) or trigger a mitigating action (if the robot is autonomous). This could augment the verification of components that can only be verified at a low level of confidence, or aid in fault finding when diagnosing failures.

\bibliographystyle{eptcs}
\bibliography{fmas}

\end{document}

%% file: acronyms.tex


\newcommand{\Genom}{GenoM}

\newacronym{ltl}{LTL}{Linear-time Temporal Logic}

\newacronym{ctl}{CTL}{Computation Tree Logic}

\newacronym{ptl}{PTL}{Probabilistic Temporal Logic}

\newacronym{pctl}{PCTL}{Probabilistic Computation Tree Logic}

\newacronym{pctl*}{PCTL*}{Probabilistic Computation Tree Logic*}

\newacronym{tctl}{TCTL}{Timed Computation Tree Logic}

\newacronym{stl}{STL}{Signal Temporal Logic}

\newacronym{ltl-x}{LTL-X}{a version of Linear Time Logic without the `next' operator}

\newacronym{mtl}{MTL}{Metric Temporal Logic}

\newacronym{iactlk-x}{IACTLK-X}{a fragment of the temporal-epistemic logic CTLK, without the `next' operator}

\newacronym{mucalc}{$\mu$-calculus}{a strict superset of other temporal logics}

\newacronym{fotl}{FOTL}{First-Order Temporal Logic}

\newacronym{bdi}{BDI}{Belief-Desire-Intetion}

\newacronym{prs}{PRS}{Procedural Reasoning System}

\newacronym{are}{ARE}{Autonomous Reasoning Engine}

\newacronym{dmars}{dMARS}{distributed Multi-Agent Reasoning System}


\newacronym{fsm}{FSM}{Finite-State Machine}

\newacronym{pfsm}{PFSM}{Probabilistic Finite-State Machine}

\newacronym{apfsm}{APFSM}{\textit{Autonomous} Probabilistic Finite-State Machine}

\newacronym{pha}{PHA}{Probabilistic Hybrid Automata}

\newacronym{fsa}{FSA}{Finite-State Automata}

\newacronym{ta}{TA}{Timed Automata}

\newacronym{gspn}{GSPN}{Generalised Stochastic Petri Net}

\newacronym{mopn}{MOPN}{Marked Ordinary Petri Net}

\newacronym{ba}{BA}{B\"{u}chi Automata}

\newacronym{dtmc}{DTMC}{Discrete-Time Markov Chain}


\newacronym{pars}{PARS}{Process Algebra for Robot Schemas}

\newacronym{fsp}{FSP}{Finite State Processes}

\newacronym{piadl}{$\pi$ADL}{$\pi$ calculus combined with ADL}

\newacronym{csp}{CSP}{Communicating Sequential Processes}

\newacronym{ccs}{CCS}{Calculus of Communicating Systems}

\newacronym{wsccs}{WSCCS}{Weighted Synchronous CCS}

\newacronym{csp|b}{CSP$\|$B}{an integrated formalism combining CSP and B}


\newacronym{fdr}{FDR}{the Failures-Divergences Refinement checker}

\newacronym{jpf}{JPF}{Java PathFinder}

\newacronym{ajpf}{AJPF}{Agent Java PathFinder}

\newacronym{mcmas}{MCMAS}{Model Checker for Multi-Agent Systems}

\newacronym{ltlmop}{LTLMoP}{Linear Temporal Logic MissiOn Planning}


\newacronym{adl}{ADL}{Architecture Description Language}

\newacronym{aadl}{AADL}{Architecture Analysis and Design Language}

\newacronym{lts}{LTS}{Labelled-Transition System}

\newacronym{uml}{UML}{Unified Modelling Language}

\newacronym{sysml}{SySML}{Systems Modelling Language}

\newacronym{robotml}{RobotML}{Robot Modelling Language}

\newacronym{dsl}{DSL}{Domain Specific Language}

\newacronym{sct}{SCT}{Supervisory Control Theory}

\newacronym{ide}{IDE}{Integrated Development Environment}

\newacronym{mde}{MDE}{Model-Driven Engineering}

\newacronym{ai}{AI}{Artificial Intelligence}

\newacronym{fdir}{FDIR}{Fault Detection, Isolation and Recovery}

\newacronym{ifm}{iFM}{Integrated Formal Methods}

\newacronym{ros}{ROS}{Robot Operating System}

\newacronym{dl}{\text{\upshape\textsf{d{\kern-0.05em}L}}}{differential dynamic logic}

\newacronym{vm}{VM}{Virtual Machine}

\newacronym{bip}{BIP}{Behaviour Interaction Priority}

\newacronym{fol}{FOL}{First-Order Logic}

\newacronym{owl}{OWL}{Web Ontology Language}

\newacronym{ria}{RIA}{Restore Invariant Approach}

\newacronym{ocl}{OCL}{Object Constraint Language}

\newacronym{sil}{SIL}{Safety Integrity Level}

%% file: dia/nodesAndVerification.tex
\begin{figure*}[t]
\centering
\scalebox{0.7}{
\begin{tikzpicture}[node distance=1em]

\node (node1) [process, xshift = -50em]{\parbox[t][][t]{3cm}{\centering{Component 1:}\\\centering{ Black-Box Implementation}}};
\node (A1)[left of = node1, xshift = -3em, yshift = 3em]{\color{black}$\mathcal{A}_1(\overline{i}_1)$};
\node (G1)[right of = node1, xshift = 3em, yshift = 3em]{\color{black}$\mathcal{G}_1(\overline{o}_1)$};

\node (node2) [process, right of = node1, xshift = 15em]{\parbox[t][][t]{3cm}{\centering Component 2:\\ Logical/Algebraic Specification}};
\node (A1)[left of = node2, xshift = -3em, yshift = 3em]{\color{black}$\mathcal{A}_2(\overline{i}_2)$};
\node (G1)[right of = node2, xshift = 3em, yshift = 3em]{\color{black}$\mathcal{G}_2(\overline{o}_2)$};

\node (node3) [process, right of = node2, xshift = 15em]{\parbox[t][][t]{3cm}{\centering{Component 3:}\\ \centering{ Model-based Specification}}};
\node (A1)[left of = node3, xshift = -3em, yshift = 3em]{\color{black}$\mathcal{A}_3(\overline{i}_3)$};
\node (G1)[right of = node3, xshift = 3em, yshift = 3em]{\color{black}$\mathcal{G}_3(\overline{o}_3)$};

\node (node4) [process, right of = node3, xshift = 15em]{\parbox[t][][t]{3cm}{\centering{Component 4:}\\\centering{ White-Box Implementation}}};
\node (A1)[left of = node4, xshift = -3em, yshift = 3em]{\color{black}$\mathcal{A}_4(\overline{i}_4)$};
\node (G1)[right of = node4, xshift = 3em, yshift = 3em]{\color{black}$\mathcal{G}_4(\overline{o}_4)$};


\node (S13) [process, below of = node1, yshift = -5em]{\parbox[t][][t]{3cm}{\centering{Software Testing}}};

\node (S14) [process, below of = node2, yshift = -5em]{\parbox[t][][t]{3cm}{\centering{Program Model Checking}}};

\node (S15) [process, below of = node3, yshift = -5em]{\parbox[t][][t]{3cm}{\centering{Theorem Proving}}};

\node (S16) [process, below of = node4, yshift = -5em]{\parbox[t][][t]{3cm}{\centering{Simulation-Based Testing}}};

\draw [arrow] (node1) -- node[above]{}(node2);
\draw [arrow] (node2) --  node[above]{}(node3);
\draw [arrow] (node3) -- node[above]{}(node4);

\draw [dashed ] (node1)  -- (S13);
\draw [dashed] (node2)  -- (S14);
\draw [dashed ] (node3)  -- (S15);
\draw [dashed ] (node4)  -- (S16);

\end{tikzpicture}

}
\caption{\small We specify the contract for each component. Here we denote the assumption/pre-condition by $\mathcal{A}(\overline{i})$ and the guarantee/post-condition by $\mathcal{G}(\overline{o})$. These contracts are  used to guide the verification approach applied to each component, denoted by dashed lines, such as software testing for a black-box implementation (Component 1). The solid arrows represent data flow. Note that we use the bar notation ($\overline{i}, \overline{o}$) to indicate a vector of variables.}
\label{fig:nodespecs}
\end{figure*}
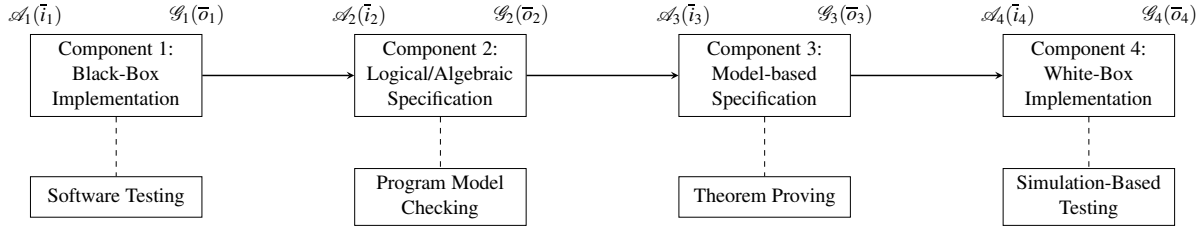

%% file: dia/robotSystem.tex
\begin{figure*}[ht]
\begin{minipage}{0.55\textwidth}
\scalebox{0.7}{
\begin{tikzpicture}[node distance=2em]

\node (vision) [process, yshift = -10em] {\parbox[t][][t]{3cm}{\centering Detection}};
\node(visionspec)[right of = vision, xshift = 6.5em]{\parbox[t][][t]{3cm}{\centering \begin{tabular}{l}$\overline{i_D}$: camera input, size  of square grid to be explored ($n$) \\ $\overline{o_D}: Grid = \{(x,y)\}, Obstacles = \{(x,y)\}, s_0 = (x,y)$\\
$\mathcal{A}_D(\overline{i_D}): n \in \mathbb{N}$\\
$\mathcal{G}_D(\overline{o_D}):\forall x,y \cdot (x,y) \in Obstacles \Rightarrow obstacle(x,y)$\\ 
\;\;\;\;\;\;\;\;\;\;\;\;\;\;$\land Grid \subseteq \mathbb{N} \times \mathbb{N} \land Obstacles \subseteq Grid \land s_0 \in Grid $\\ 
\;\;\;\;\;\;\;\;\;\;\;\;\;\;$\land s_0 \notin Obstacles \land  \forall x,y \cdot (x,y) \in Grid \Rightarrow x<n \land y<n$\end{tabular}}};

\node (planner) [process, below of=vision, yshift=-6em]{\parbox[t][][t]{3cm}{\centering Planner}};
\node(plannerspec)[right of = planner, xshift = 6.5em]{\parbox[t][][t]{3cm}{\centering \begin{tabular}{l}\textbf{$\overline{i_P}$:} $\overline{o_D}$\\ \textbf{$\overline{o_P}$:} $PlanSet = \{\{(x, y)\}\}$\\ 
$\mathcal{A}_P(\overline{i_P}): \mathcal{G}_D(\overline{o_D})$\\
$\mathcal{G}_P(\overline{o_P}): \forall p \cdot p \in PlanSet \Rightarrow  p \subseteq Grid\setminus Obstacles \land s_0 \in p$\\
\;\;\;\;\;\;\;\;\;\;\;\;\;\;$\land \exists p_0 \cdot p_0 \in p \land adjacent(s_0,p_0)$\\
\;\;\;\;\;\;\;\;\;\;\;\;\;\;$\forall p_1\cdot p_1 \in p \land p_1 \neq s0 \Rightarrow$\\
\;\;\;\;\;\;\;\;\;\;\;\;\;\;$(\exists r,s \cdot r\in p \land s\in p \land adjacent(r,p_1) \land adjacent(p_1,s))$
\end{tabular}}};

\node (agent) [process,  below of = planner, yshift = -6em]{\parbox[t][][t]{3cm}{\centering Agent}};
\node(agentspec)[right of = agent, xshift = 6.5em]{\parbox[t][][t]{3cm}{\centering \begin{tabular}{l}\textbf{$\overline{i_A}$:} $\overline{o_P}$ \\ \textbf{$\overline{o_A}$:} $plan = \{(x, y)\}$\\
$\mathcal{A}_A(\overline{i_A}): \mathcal{G}_P(\overline{o_P})$\\
$\mathcal{G}_A(\overline{o_A}): plan \in PlanSet \land \forall q \cdot q \in PlanSet \Rightarrow |plan| \leq |q|$\end{tabular}}};

\draw [arrow] (vision) -- node[right]{$\overline{o_D}$}(planner);
\draw [arrow] (planner) --node[right]{$\overline{o_P}$}(agent);

\end{tikzpicture}}
\end{minipage}
\hfill\vline\hspace{11pt}
\begin{minipage}{0.34\textwidth}
\begin{programsc}
\EVT{Planner}{ordinary}
 \bkw{any} $p$
 \bkw{where}
  \bLabel{grd1}{goal\neq{}s_0 \land{}goal\in{}p}
  \bLabel{grd2}{p\subseteq{}Grid\setminus{}Obstacles \land s_0\in{}p}
  \bLabel{grd4}{\exists{}p_0\cdot{}p_0\in{}p\land{}adjacent(s_0\mapsto{}p_0)}
  \bLabel{grd6}{\forall{}p_1\cdot{}p_1\in{}p\land{}p_1\neq{}s_0\land{}p_1\neq{}goal}\\   $\Rightarrow{}(\exists{}r,s\cdot{}r\in{}p\land{}s\in{}p\land{} adjacent(r\mapsto{}p_1)\land{}adjacent(p_1\mapsto{}s))$
  \bLabel{grd5}{\exists{}p_2\cdot{}p_2\in{}p\land{}adjacent(p_2\mapsto{}goal)}
 \bkw{then}
  \bLabel{act1}{PlanSet := PlanSet \cup \{p\}}
\end{programsc}
\end{minipage}
\caption{\small An overview of the robotic system to be verified. On the far left, each rectangle represents a component, and each arrow represents data flow between components. To the right of each individual component, we summarise the $\overline{i}$s, $\overline{o}$s, and their respective FOL contract. On the far right, we provide an Event-B event specification corresponding to the \emph{Planner}. Here, `$\mapsto$' denotes tuples.}
\label{fig:system}

\end{figure*}
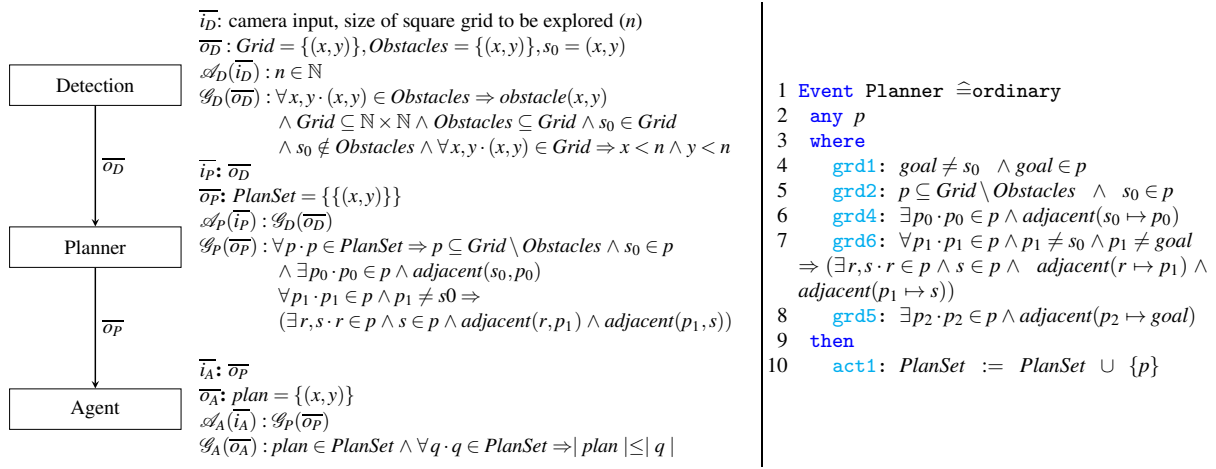